\documentclass[12pt]{iopart}

\usepackage{graphicx}
\def\ba{\begin{eqnarray}}
\def\ea{\end{eqnarray}}

\begin{document}

\title[Momentum constraint relaxation]{Momentum constraint relaxation}

\author{Pedro Marronetti}

\address{Department of Physics,
Florida Atlantic University, Boca Raton, FL 33431}

\begin{abstract}
Full relativistic simulations in three dimensions invariably
develop runaway modes that grow exponentially and are accompanied by
violations of the Hamiltonian and momentum constraints.
Recently, we introduced a numerical method (Hamiltonian relaxation)
that greatly reduces the Hamiltonian constraint violation and helps
improve the quality of the numerical model. We present here a
method that controls the violation of the momentum constraint. The
method is based on the addition of a longitudinal component to the
traceless extrinsic curvature $\tilde{A}_{ij}$, generated by a vector
potential $w_i$, as outlined by York. The components of $w_i$ are relaxed
to solve approximately the momentum constraint equations, pushing slowly
the evolution toward the space of solutions of the constraint equations.
We test this method with simulations of binary neutron stars in
circular orbits and show that effectively controls the growth of the
aforementioned violations. We also show that a full numerical enforcement
of the constraints, as opposed to the gentle correction of the momentum
relaxation scheme, results in the development of instabilities that
stop the runs shortly.

\end{abstract}


\ead{pmarrone@physics.fau.edu}

\maketitle


\section{Introduction}
\label{intro}

For most of the past decade, the main theoretical thrust
in gravitational research has been directed toward
obtaining stable and accurate numerical models of compact-object
binary systems. One of the most difficult problems to tackle has
been the control of exponentially growing instabilities that degrade
the quality of any simulation and, eventually, terminate it.
General relativistic numerical simulations of BNS systems have made 
significant progress in the past years \cite{Shibata:1999wm,Shibata:2002jb,
Duez:2002bn,Shibata:2003ga,Miller:2003vc,Marronetti:2003hx,
Kawamura:2003hu}, in particular with the addition of advanced numerical 
techniques such as AMR \cite{Evans:2005mt}.

The numerical formulations employed to simulate astrophysical
systems can be divided in unconstrained and constrained methods.
Unconstrained formulations like ADM \cite{adm62}
and BSSN \cite{Shibata:1995,Baumgarte:1998te} simply evolve the
gravitational fields without any attempts at controlling the violation
of the time independent Hamiltonian and momentum constraints; these
equations are merely monitored to gauge the accuracy of the model.
Constrained formalisms, on the other hand, enforce
the satisfaction of the  constraint equations at either the analytical
level (i.e., ingraining them in some way in the time evolution equations)
or at the numerical level (i.e., regularly solving the constraints
during the evolution).
While constrained methods have been applied mostly to problems with
spherical \cite{Choptuik:1992jv} and axial symmetry
\cite{Abrahams:1994ge,Choptuik:2003as,Choptuik:2003ac},
there have been recent applications to
three-dimensional scenarios \cite{Bonazzola:2003dm,Anderson:2003dz}.

We propose in this and a companion article \cite{Marronetti:2005bz}
(from now on, Paper I) an alternative type of evolution in which the
Hamiltonian and momentum constraints are only approximately solved at
every time step, gently steering the evolution toward the space of
solutions of these equations without completely forcing their
numerical satisfaction. In Paper I we described a method that controls the
Hamiltonian constraint violation ({\it Hamiltonian relaxation} or HR).
In this paper we introduce a complementary scheme that reduces the
momentum constraint residuals ({\it momentum relaxation} or MR).
The constraint relaxation techniques utilize the conformal
decomposition of the spatial metric and extrinsic curvature presented
by York \cite{y73,y79}, which has traditionally been used to solve the
initial value problem for binary systems
~\cite{Matzner:1998pt,Marronetti:2000rw,Marronetti:2000yk,Pfeiffer:2002xz,Bonning:2003im}.
In this decomposition, a conformal factor $\psi$ factored out of the spatial
metric and a longitudinal addition to the extrinsic curvature generated
from a vector potential $w_i$ are used to satisfy the Hamiltonian and
the three components of the momentum constraint respectively. HR
drives $\psi$ to the solution space of the Hamiltonian constraint
by means of a parabolic equation for the conformal factor,
in the spirit of the K-Driver \cite{betal96} used for the lapse.
The MR method described in this paper uses $w_i$ to
push the simulation toward the space of solutions of the momentum
constraint. In both cases, a full relaxation of $\psi$ and $w_i$
would lead to the numerical solution of the constraints. However, the
stability of the relaxation methods relies on gently updating these
fields during the evolution. We show in \ref{appendix_CT} that a full
relaxation scheme becomes unstable rather quickly when used in combination
with BSSN.

We tested the new algorithms by simulating binary neutron stars (BNS) in
circular orbits and show that the use of these techniques results in a
notable improvement of the overall quality of the simulation. HR not only
suppresses the Hamiltonian constraint violation but also contributes to
a more stable behavior in the total angular momentum of the system.
MR contribution is mostly confined to quenching the momentum constraint
violation. The BNS simulation runs for about two orbits before stopping.
One of the main reasons for the degradation of the simulation quality is
the use of a shift vector frozen to its initial value
($\beta$-freeze). This choice, while very convenient when testing new
algorithms, becomes inadequate as soon as the stars move appreciably
from their initial position. Given that our simulations are performed
in the frame that rotates with the binary, this occurs rather late in
the run. The second cause is related to inappropriate boundary conditions
for the rest of the gravitational fields. The use of radiation
(Sommerfeld) conditions in rotating frames becomes troublesome for
large grids in rotating frames. These problems will be addressed in
future work.

Section \ref{EE} describes in detail the momentum constraint relaxation
method and its numerical implementation. Section \ref{results} presents
simulations of BNS in circular orbits and compares the results obtained
with and without the relaxation techniques. Section \ref{conclusions}
summarizes our results and \ref{appendix_CT} presents the convergence
tests.


\section{Equations for the Gravitational and Hydrodynamical Fields}
\label{EE}

\subsection{Time Evolution Equations}
\label{TEE}

We use geometrized units ($G=c=1$) and the Greek (Latin) indices run
from 0 to 3 (1 to 3). In the standard ``3+1" form, the metric is written
as
\ba
\rmd s^2 = -\alpha^2 \rmd t^2 + \gamma_{ij}(\rmd x^i+\beta^i \rmd t)
(\rmd x^j+\beta^j \rmd t)~,\nonumber
\ea
where $\alpha$, $\beta^i$, and $\gamma_{ij}$ are the lapse function,
shift vector, and spatial metric tensor, respectively.

The ADM formulation \cite{adm62} splits the Einstein's field equations
\ba G_{\mu\nu} = 8 \pi T_{\mu\nu} \nonumber \ea
into a set of time-independent elliptic equations
\ba R - K_{ij}K^{ij} + K^2 = 16\pi\rho~, \label{HC} \\
D_jK^j_i - D_iK = 8\pi S_i~, \label{MC} \ea
known as the Hamiltonian and momentum constraints, plus a set of
time-dependent hyperbolic equations:
\ba
\fl (\partial_t - {\mathcal{L}}_{\beta})\gamma_{ij} = -2 \alpha K_{ij}~,\\
\fl (\partial_t - {\mathcal{L}}_{\beta}) K_{ij}=
        - D_iD_j\alpha + \alpha \{ R_{ij}  -  2K_{il}K^l{}_j + K K_{ij}
        - 8\pi [S_{ij}  +  \frac{1}{ 2} \gamma_{ij} (\rho-S)]\}~.
\label{ADM}
\ea
The latter set provides the evolution in time of the spatial
metric $\gamma_{ij}$ and the extrinsic curvature $K_{ij}$. The symbol
$D_i$ represents the covariant gradient with respect to the tensor
$\gamma_{ij}$. The fields $\rho$, $S$, and $S_{ij}$ are derived from the
matter fields by splitting the stress-energy tensor $T_{\mu\nu}$ in components
parallel and perpendicular to the normal of the spatial hypersurface
$n^\alpha$ \cite{Marronetti:2005bz}.

Following York \cite{y73,y79}, we can decompose the tensors $\gamma_{ij}$ and $K_{ij}$
as
\ba
\gamma_{ij} = \psi^4 ~\tilde{\gamma}_{ij}~, \nonumber\\
K_{ij} = \psi^4 ~\bigl(\tilde A_{ij} + \frac{1}{3}
\tilde\gamma_{ij}K\bigr)~, \nonumber
\ea
Where the fields $\psi$, $\tilde{\gamma}_{ij}$,
$\tilde A_{ij}$, and $K$, are known as the conformal factor, the conformal metric,
the conformal traceless extrinsic curvature, and the trace of the extrinsic
curvature respectively. We can write the Hamiltonian and momentum constraints
using the new variables as
\ba \label{HC2}
\tilde\gamma^{ij}\tilde D_i\tilde D_j \psi
           - \frac{\psi}{8}\tilde R
       + \frac{\psi^5}{8}\tilde A_{ij}\tilde A^{ij}
           -  \frac{\psi^5}{12}K^2
       +  2\pi \psi^5 \rho  =  0, \\
\label{MC2}
\tilde D^j(\psi^6 \tilde A_{ji})- \frac{2}{3} \psi^6 \tilde D_i K
  - 8\pi \psi^6 S_i
  = 0~.
\ea
We define the Hamiltonian constraint residual $\mathcal{H}$ and momentum
constraint residual $\mathcal{M}_i$ as the l.h.s. of equations (\ref{HC2})
and (\ref{MC2}) respectively.

The BSSN formulation \cite{Shibata:1995,Baumgarte:1998te} provides a set of
evolution equations for the fields $\psi$, $\tilde{\gamma}_{ij}$, $K$, and
$\tilde A_{ij}$
\ba 
\fl (\partial_t - {\mathcal{L}}_{\beta}) \ln(\psi)
= - \frac{1}{6}\alpha K ~, \label{cf_dot_BSSN}\\
\fl (\partial_t - {\mathcal{L}}_{\beta})\tilde\gamma_{ij}
                = -2\alpha\tilde A_{ij}~, \label{g_dot_BSSN} \\
\fl (\partial_t - {\mathcal{L}}_{\beta})K
                = -\gamma^{ij}D_jD_i\alpha + \frac{1}{3}\alpha K^2
                  + \alpha \tilde A_{ij}\tilde A^{ij}
                    + 4\pi\alpha (\rho + S) ~,\label{K_dot_BSSN} \\
\fl (\partial_t - {\mathcal{L}}_{\beta})\tilde A_{ij}
                = \psi^{-4} [-D_iD_j\alpha
                    + \alpha(R_{ij}-8\pi S_{ij})]^{TF}
                  + \alpha(K\tilde A_{ij} - 2\tilde A_{il}\tilde A^l{}_j)~, 
		  \label{A_dot_BSSN}
\ea
where the superscript $TF$ indicates the trace-free part of the tensor. These
fields are complemented with the variable known as the conformal connection
\ba
\tilde\Gamma^i  \equiv -\tilde\gamma^{ij}{}_{,j},\nonumber
\ea
and its corresponding evolution equation
\ba
\fl \partial_t\tilde\Gamma^i &= \partial_j(2\alpha\tilde A^{ij}
                + {\mathcal{L}}_{\beta}\tilde\gamma^{ij}) \nonumber \\
\fl      &=  \tilde\gamma^{jk}\beta^i{}_{,jk}
            + \frac{1}{3}\tilde\gamma^{ij}\beta^k{}_{,kj}
            - \tilde\Gamma^j\beta^i{}_{,j}
        + \frac{2}{3}\tilde\Gamma^i\beta^j{}_{,j}
            + \beta^j\tilde\Gamma^i{}_{,j}
        - 2\tilde A^{ij}\partial_j\alpha \nonumber \\
\fl       & - 2\alpha ~(\frac{2}{3} \tilde\gamma^{ij}K_{,j}
            - 6\tilde A^{ij} ~[\ln(\psi)]_{,j}
            - \tilde\Gamma^i{}_{jk}\tilde A^{jk}
        + 8\pi\tilde\gamma^{ij}S_j)~.\label{Gamma_dot_BSSN}
\ea

ADM and BSSN are, in their original forms, unconstrained formulations.
The Hamiltonian and momentum constraints are not enforced throughout
the simulation but only monitored for quality control. Anderson \& Matzner
\cite{Anderson:2003dz} presented a constrained variation of the ADM
formalism where the constraints are solved numerically at every time step.
They use equation (\ref{HC2}) to solve numerically for the conformal factor $\psi$,
while the momentum constraint equations (\ref{MC2}) are satisfied by the addition
of a longitudinal component to the traceless conformal extrinsic curvature
$\tilde{A}^{ij}$ as originally outlined by York \cite{y79}. We introduced in
Paper I the Hamiltonian relaxation technique that draws on the same idea of
controlling the Hamiltonian constraint violating modes by adjusting
$\psi$. HR uses the conformal factor to solve {\it approximately}
the Hamiltonian constraint (\ref{HC2}), bypassing altogether the
corresponding evolution equation (\ref{cf_dot_BSSN}) for $\psi$ .
Instead of solving equation (\ref{HC2}), HR composes an alternative parabolic equation
which relaxes $\psi$ towards a solution of the Hamiltonian constraint through an
iterative scheme. The equation used in Paper I was of the form
\ba
\partial_{t} \psi = \epsilon_H ~(\partial_{t} \mathcal{H} + \eta_H ~\mathcal{H})~,
\label{psi_dot} \ea
where $\epsilon_H$ and $\eta_H$ are fine-tuning parameters. This technique
efficiently quenches the development of Hamiltonian constraint violation
instabilities, improving the overall quality of BNS simulations.

In this paper we present the momentum constraint relaxation method, which suppresses
the development of momentum constraint violation instabilities and works as a
complement to HR. MR is based on correcting the traceless conformal
extrinsic curvature $\tilde{A}_{ij}$ with a longitudinal component
$\tilde{(lw)}_{ij}$ \cite{y79}. The operator $\tilde{(lw)}_{ij}$ is defined as
\ba \label{lw}
\tilde{(lw)}_{ij} \equiv \tilde{D}_i w_j + \tilde{D}_j w_i
- \frac{2}{3}~ \gamma_{ij}~ \tilde{D}^n w_n  ~,
\ea
and $w_i$ is known as the vector potential. The momentum constraint
equations (\ref{MC2}) become now a function of $w_i$
\ba \label{MC3}
\mathcal{M}_i = \tilde{\gamma}^{kj} \left[ ~\tilde{D}_k \tilde{(lw)}_{ij}
    + 6~ ~\tilde{(lw)}_{ij} ~\tilde{D}_k \ln \psi \right]
    + \rho_i = 0~,
\ea
where $\rho_i$ collects all terms independent of the vector potential
\ba
\rho_i \equiv \tilde{\gamma}^{kj} \left[ ~\tilde{D}_k \tilde{A}_{ij}
    + 6~\tilde{A}_{ij} ~\tilde{D}_k \ln \psi \right]
    - \frac{2}{3}~\tilde{D}_i K - 8 ~\pi ~S_i~.
\ea

Equations (\ref{MC3}) form a set of coupled elliptic equations for the components
of the vector potential $w_i$ that can be solved numerically using a number of
different algorithms. Note a formal difference between HR and MR: HR replaces the 
BSSN evolution equation for the conformal factor (\ref{cf_dot_BSSN}) with a new
equation (\ref{psi_dot}), while MR adds three new auxiliary fields, the $w_i$
components, with their corresponding equations (\ref{MC3}).

The components of the vector potential $w_i$ and the conformal factor $\psi$
are added to the list of dynamical fields evolved using BSSN: $\tilde{\gamma}_{ij}$,
$\tilde A_{ij}$, $K$, and $\tilde{\Gamma}^i$. It is important to note
that the fields $\tilde A_{ij}$ are {\it not} updated using the $w_i$ new values. 
Instead, the $w_i$ are kept as dynamical variables that {\it evolve}
from one time step to the next. The r.h.s. of the BSSN evolution equations
are modified to include the contribution of $w_i$; in equations
(\ref{g_dot_BSSN}-\ref{Gamma_dot_BSSN}), 
$\tilde A_{ij}$ is replaced by $\tilde A_{ij}+\tilde{(lw)}_{ij}$. This facilitates
the implementation of $w_i$ boundary conditions, as explained in section
\ref{BCs}.

The work presented in this paper is based on a Successive Under Relaxation algorithm 
(SUR) similar to the one described in \cite{NumRec}. This algorithm updates the 
value of $w_i$ at each relaxation step $r$ with a correction term
\ba
w_i^{(r)} = w_i^{(r-1)} - \omega_M ~ \Delta w_i^{(r-1)}~~,
\ea
where $\omega_M$ is a relaxation parameter such that $0<\omega_M<2$
\cite{NumRec}. When $\omega_M > 1$ (i.e.,
{\it Successive Over Relaxation} or SOR) the method converges faster to
the numerical solution than for the case $\omega_M < 1$. However, a key concept
behind momentum relaxation is to use the vector potential $w_i$ to gradually
project the simulation onto the space of solutions of the momentum constraint
equations. Since we are interested in gentle updates of $w_i$, we use values
of $\omega_M < 1$. SUR also allows for a rather straightforward implementation
of the boundary conditions for the vector potential described in section
\ref{BCs}.

Their values are updated at each one of the steps of the Iterative Crank-Nicholson 
(ICN) method (one Predictor and two Corrector steps) used for time integration. The 
implementation of momentum relaxation, which in this paper is used in combination 
with HR, is as follows :\\

1) Initial update of $w_i$ : After the initial data set corresponding to a BNS
system in circular orbit is read, $w_i$ undergoes a relatively large number
of relaxation steps (typically about 90), starting from the initial guess
$w^i=0$. The components are relaxed sequentially in the order $w^z$, $w^y$, and
$w^x$. Given that the three equations are coupled in the components of $w_i$, the
last component to be relaxed experiences the smoother update. We choose the
$x$ component last because the violation of this component of the momentum
constraint is the largest due to our choice of coordinate axis (binary aligned
along the $y$ axis and $z=0$ the orbital plane).

2) At the Predictor stage of the ICN integration, we perform the update of the
fields in the following order:

i) Update of the gravitational fields $\tilde{\gamma}_{ij}$, $\tilde A_{ij}$,
$K$, and $\tilde{\Gamma}^i$.

ii) Update of the conformal factor $\psi$ using HR (Paper I).

iii) Update of the vector potential components $w^z$, $w^y$, and $w^x$.
The sources of the elliptic equations are calculated using the values of the fields
corresponding to the previous time step. Each equation undergoes typically from
10 to 20 relaxation steps. The boundary conditions for $w_i$ are updated
right after each relaxation step (section \ref{BCs}).

iv) Update of the hydrodynamical fields, the lapse function, and the shift vector.\\

The boundary conditions are updated at the same time as their corresponding fields.
The same steps are followed in the first Corrector (second Corrector)
stage, but replacing the field values from the previous time step with
the corresponding updates generated in the Predictor (first Corrector)
step.


\subsection{Lapse and Shift Equations}

We use the same lapse function and shift vectors employed in
the development of HR (Paper I):
the K-Driver algorithm \cite{betal96} for $\alpha$ and the shift vector remains
unchanged from its initial value ($\beta$-freezing). All the simulations
presented in this paper were performed in the frame that rotates with the
binary (corotating), given its superior stability properties over inertial frames
\cite{Duez:2002bn, Marronetti:2003hx, Swesty:1999ke}. We refer the reader to
Paper I for the details of the numerical implementation as well as the boundary
conditions.


\subsection{Boundary Conditions}

We adopt Sommerfeld (radiative) boundary conditions for the conformal
metric $\tilde{\gamma}_{ij}$, the traceless part of the extrinsic
curvature $\tilde{A}_{ij}$, and the trace of the extrinsic curvature
$K$, and we set $\tilde{\Gamma}^i=0$ at the boundaries \cite{Duez:2002bn}.
The boundary conditions for the conformal factor $\psi$ are such
that enforce the satisfaction of $\mathcal{H}=0$ in its
finite-difference form. They are essential for the control of the
Hamiltonian constraint violating modes coming from the grid
boundaries.

\subsubsection{Boundary Conditions for the Vector Potential $w_i$}
\label{BCs}

The boundary conditions for the vector potential $w_i$ are inspired
in their conformal factor counterparts and also aim at controlling
as effectively as possible the incoming constraint violating modes.
The boundary conditions for the vector potential $w_i$ are such that
they enforce the satisfaction to round-off error of the momentum
constraint at the grid points next to the boundaries.
Equation (\ref{MC3}) can be expanded into
\ba \label{MC4}
\fl    \tilde{\gamma}^{kj} \left[ ~\tilde{D}_k \tilde{(lw)}_{ij}
    + 6~ \tilde{(lw)}_{ij} ~\tilde{D}_k \ln \psi \right]+\rho_i ~= \nonumber\\
\fl    \tilde{\gamma}^{kj} \left[ ~\tilde{D}_k \tilde{D}_i w_j
    + ~\tilde{D}_k \tilde{D}_j w_i \right]
    - \frac{2}{3} ~\tilde{D}_i ~\tilde{D}^n w_n +
      6~ \tilde{\gamma}^{kj} ~\partial_k (\ln \psi)
      \left[ \tilde{D}_i w_j + \tilde{D}_j w_i
      - \frac{2}{3} \tilde{\gamma}_{ij} ~\tilde{D}^n w_n \right] \nonumber\\
      +\rho_i = 0~,
\ea
where the first and second covariant derivatives of $w_i$ are,
as usual, functions of the partial derivatives with respect to the
spatial coordinates. These partial derivatives of the vector potential
are approximated in our finite-difference scheme by second order
stencils of the form
\ba \label{FD}
\partial_x w_i^n & \sim &
    \frac{w_{i(N,j,k)}^n-w_{i(N-2,j,k)}^n}{2\Delta x}~, \nonumber \\
\partial_x \partial_x w_i^n & \sim &
    \frac{w_{i(N,j,k)}^n-2 w_{i(N-1,j,k)}^n
    + w_{i(N-2,j,k)}^n}{(\Delta x)^2}~, \nonumber\\
\partial_x \partial_y w_i^n & \sim & \frac{w_{i(N,j+1,k)}^n- w_{i(N-2,j+1,k)}^n
-w_{i(N,j-1,k)}^n+w_{i(N-2,j-1,k)}^n}{4~\Delta x \Delta y}~, \nonumber\\
\rm{etc,}
\ea
where $n$ is the relaxation iteration step, $\Delta x$ and $\Delta y$
the grid spacing along the $x$ and $y$ axis, respectively. The above formulae
have been written for the grid points next to the boundary
$x = x_{max}$ (i.e.; the points with indices $(N-1,j,k)$). When
we replace the differential operators of equation (\ref{MC4}) with the
finite-difference formulae (\ref{FD}) and solve for the boundary values
of $w_i$ at the boundary $x = x_{max}$, we obtain
\ba \label{wi_BC}
w_{i(N,j,k)}^n = F_{i(N,j,k)} ~,
\ea
where $F_{i(N,j,k)}$ are algebraic expressions which depend on $w_{i(N-1,j,k)}^n$,
$w_{i(N,j-1,k)}^n$, $w_{i(N,j+1,k)}^n$, $w_{i(N,j,k-1)}^n$, and
$w_{i(N,j,k+1)}^n$. The linear system of equations (\ref{wi_BC}) is quite
large even for modest-size grids and thus very time consuming to solve for.
To expedite the process, we evaluate the equations (\ref{wi_BC})
using the newly calculated values at the boundaries as soon as they
are ready (i.e., $w_{i(N-1,j,k)}^n$, $w_{i(N,j-1,k)}^n$, and
$w_{i(N,j,k-1)}^n$) and use the previous iteration
values for the remainder (i.e., $w_{i(N,j+1,k)}^{n-1}$ and $w_{i(N,j,k+1)}^{n-1}$).
This approximation is justified by the fact that these boundaries are
evaluated many times in every time step.

Similar to the case of the boundary conditions for the conformal factor,
a problem arises at the edges of the grid, where the expression for
the boundary value $w_{i(N,N,k)}^n$ is now a function of the inverse
of the spatial conformal metric $\tilde{\gamma}_{xy}$, which is
null at the initial time step. For these points we use bilinear
interpolation until the value of $\tilde{\gamma}_{xy}$ crosses some
predetermined value which, for the simulations of this paper, is
$10^{-6}$.


\subsection{Hydrodynamical Formulation and Initial Data sets}

The numerical simulation of neutron star spacetimes requires
algorithms for the evolution of the matter fields. The fluid in
the stars is described by a perfect-fluid stress-energy tensor
and a polytropic equation of state with constant $\Gamma=2$ is
assumed. We use the hydrodynamical methods described in Paper I.
The initial state of the BNS in circular orbit is obtained
using the Wilson-Mathews Conformally Flat Condition (CFC) approach
\cite{Wilson:1995ty, Wilson:1996ty}, using the elliptic solver
described in \cite{Marronetti:2003gk}.

We tested the numerical implementation of the momentum relaxation
method using short trial-and-error runs that consisted essentially
in BNS simulations performed in small low-resolution grids.
To simplify even further these runs, they were based on an initial
data set corresponding to a corotating (tidally-locked)
binary. The satisfactory performance of these short runs was
followed by longer runs based on larger grids and higher resolution.
These long runs were also based on more realistic irrotational (zero-spin)
BNS systems. The details of the initial data sets used in these runs
is given in table I of Paper I. The number of points in the short
run grid is about 73 times smaller than that of the long run
simulations, making it possible to simulate a BNS orbital period
in a couple of hours on a typical single-processor workstation.
The long runs were performed using the IA-64 Linux Beowulf
cluster {\it Mercury} at NCSA.

Our simulations are performed in Cartesian grids and use
finite-difference second order operators within grids that have
uniform and identical spacing along each axis. We work in the reference
frame that rotates with the binary, and the stars are aligned with
the $y$ axis.


\section{Results}
\label{results}

\subsection{Short Trial-and-Error Runs}
\label{short_runs}

\begin{figure}
\begin{center}
\caption{ \underline{Short Runs}: Evolution of the three components of
the momentum constraint residual during the first steps of the
time evolution. The upper (lower) plot corresponds to a run without
(with) momentum relaxation. The thick line is the initial residual.
The curves are plotted following the line with coordinates $(0,y,0)$,
that runs through the center of the star. The companion star, located
on the negative $y$ hemisphere, is not shown.}
\label{Mi_MR_Evol}
    \includegraphics[width=3.0in]{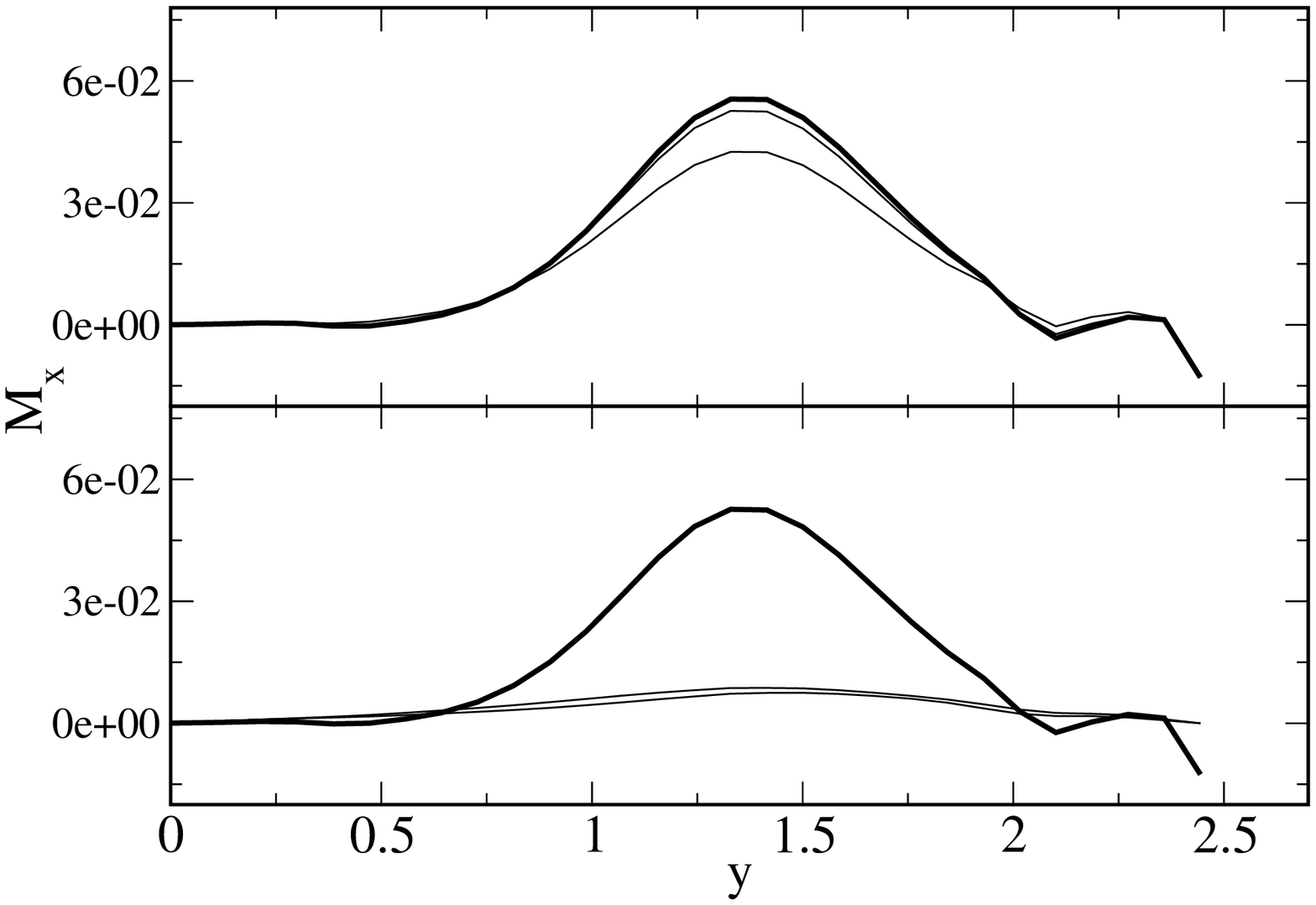}
    \includegraphics[width=3.0in]{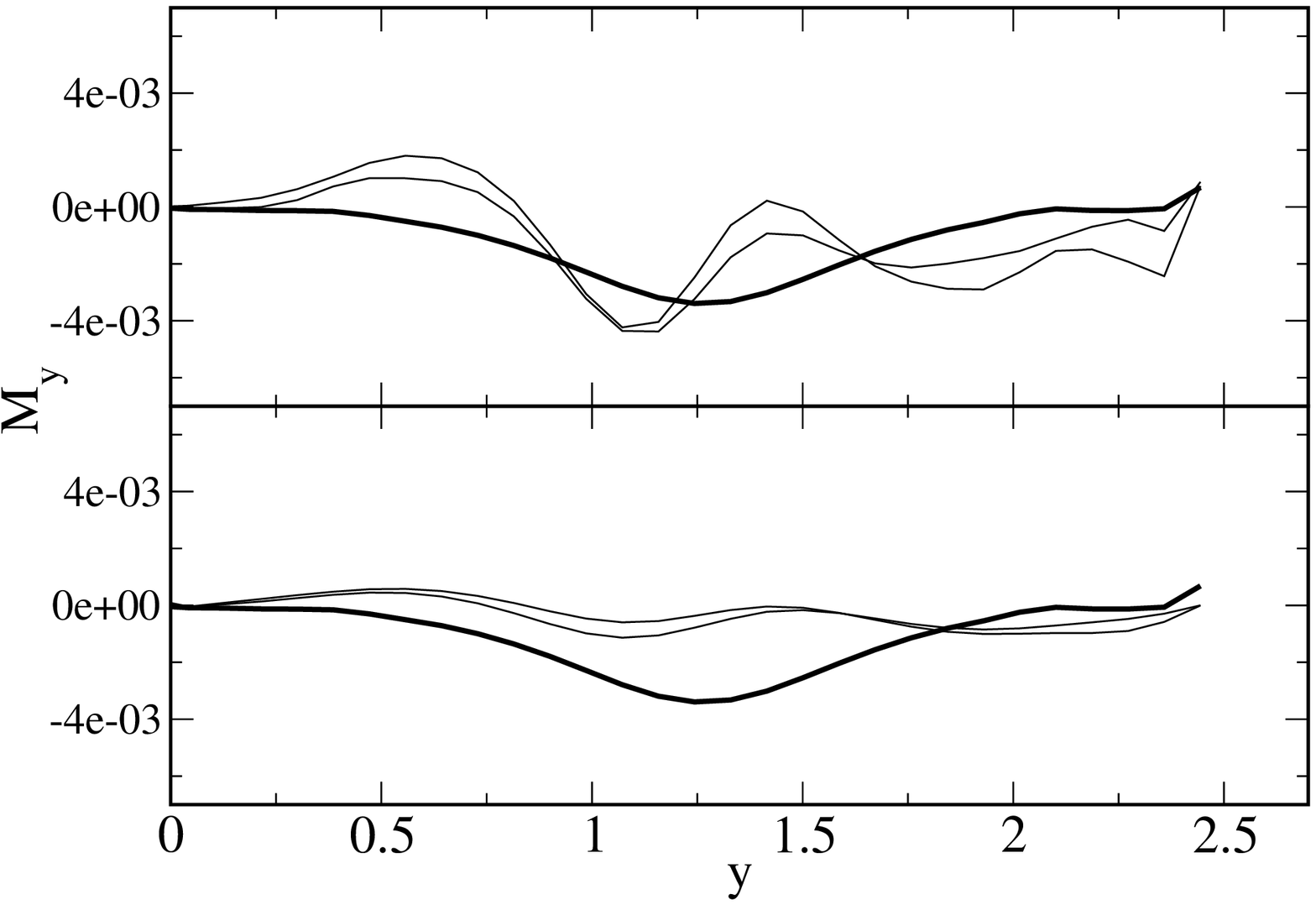}
    \includegraphics[width=3.0in]{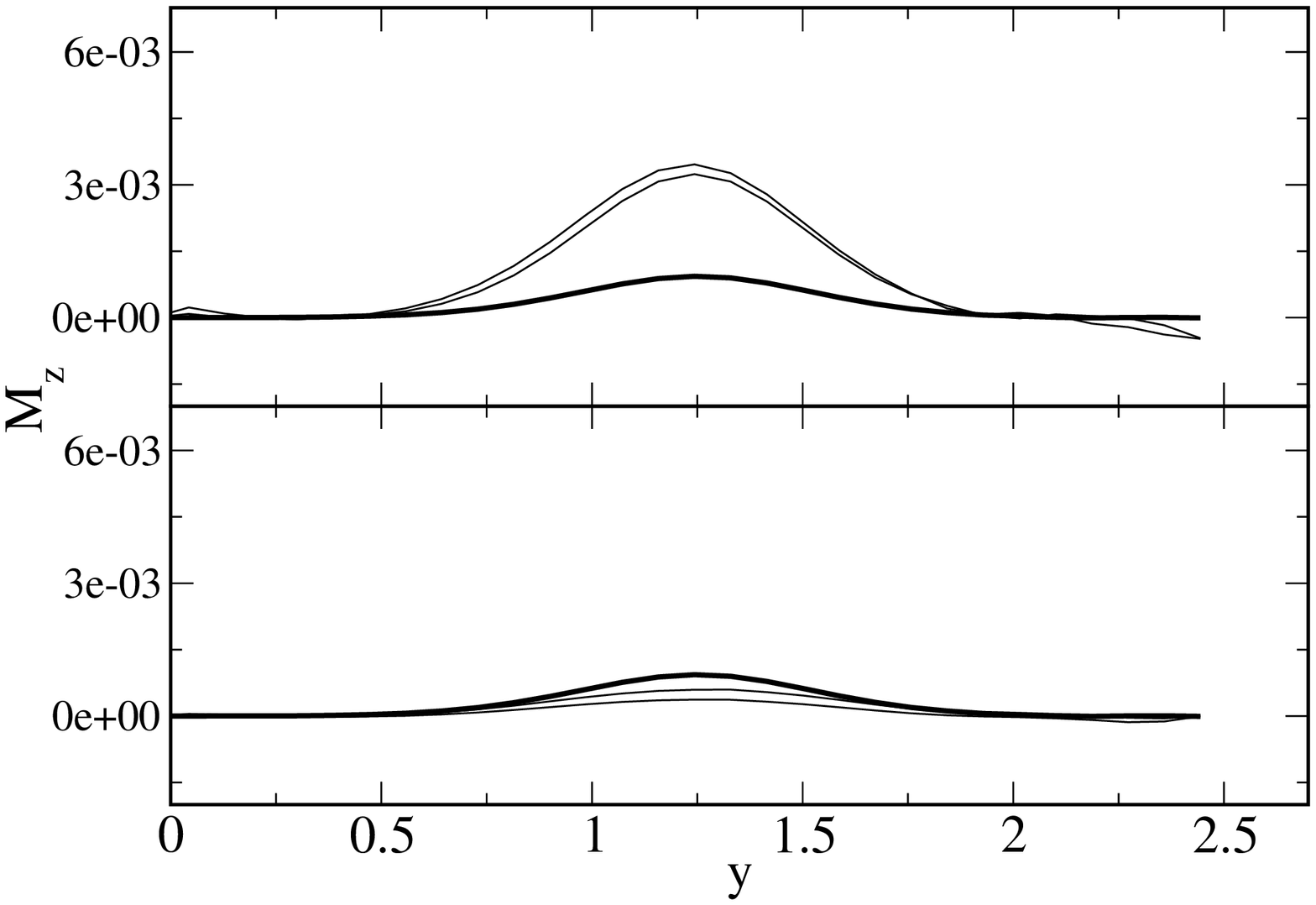}
\end{center}
\end{figure}

The free parameters of MR are the relaxation parameter $\omega_M$
and the number of relaxation steps. Their corresponding values were
determined empirically and set to $\omega_M=0.1$ with 20 relaxation steps
for the short runs. Note that $\psi$ and $w_i$ are relaxed
three times at every time step (i.e., in the ICN predictor and
the two corrector stages).

Following the HR method used for the short runs of Paper I, we use
the parameter values $\epsilon_H=0.0001$ and $\eta_H = 70.0$, and the
relaxation of $\psi$ is performed until the $L_2$ norm of the
Hamiltonian residual is smaller than the one at the previous time step
for up to a maximum of 25 iterations.

For comparison, we show in this and the next section curves
corresponding to runs performed using only the BSSN formulation
as described in \cite{Duez:2002bn}, runs using only HR, and
runs using both HR and MR. The BSSN runs differ from the HR and HR+MR
runs in two aspects. One is that the BSSN runs use the $\Gamma$-Driver
\cite{ab01,Duez:2002bn} while the HR only and HR+MR runs use the
$\beta$-freeze condition for the shift vector. The other difference is
that we use fall-off Robin-like boundary conditions for the lapse
function for the BSSN runs and frozen values for the HR and HR+MR runs
(see Paper I). All the runs (short and long) have Courant factors of
$0.46$. We also conserve the K-Driver and $\Gamma$-Driver (BSSN runs only)
parameters, which  are set to $\epsilon_\alpha=0.125$, $\eta_\alpha=0.1$,
$\epsilon_\beta=0.0005$, $\eta_\beta=0.2$, respectively. The K-Driver
($\Gamma$-Driver) relaxation is iterated 5 (10) times.

\begin{figure}
\begin{center}
\caption{ \underline{Short Runs}: Total angular momentum $J$ as a
function of time, given as fraction of the orbital period $P$. $J$
is normalized to its initial value $J_0$. The lines correspond to runs
using BSSN (dotted), HR only (thin solid), and HR + MR (thick solid).}
\label{J_Short_Runs}
    \includegraphics[width=5.0in]{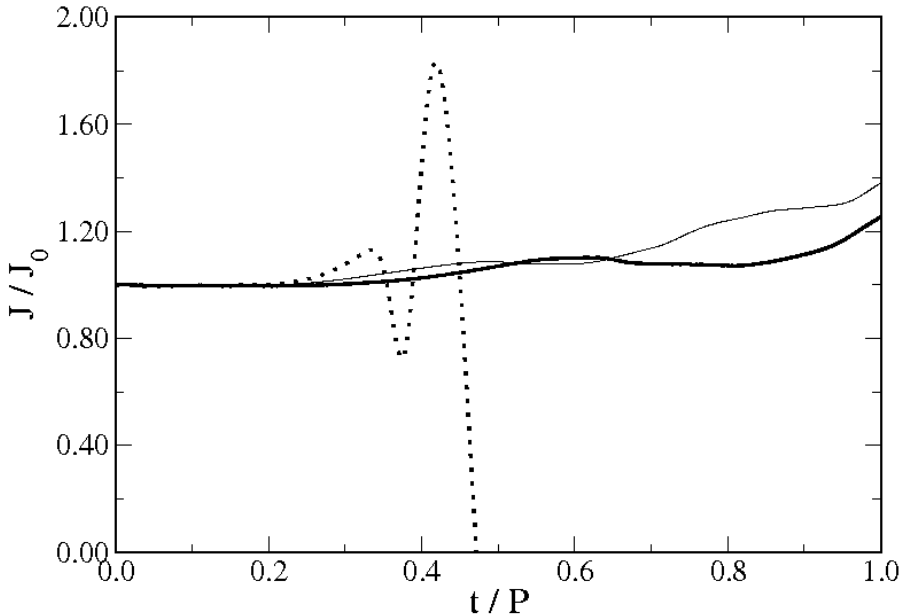}
\end{center}
\end{figure}

Figure \ref{Mi_MR_Evol} presents the evolution of the momentum
constraint residuals $\mathcal{M}_i$ for the first three time steps
of a BNS simulation without (with) MR in the upper (lower) part of the plots.
The momentum constraint violation is plotted along the binary axis ($y$).
Only one star is present in the grid (the companion star is in the $y<0$
hemisphere) and its center is at $y \simeq 1.4$. The thick lines correspond
in all cases to the initial data. The reduction of the momentum constraint
violation caused by MR can be seen both in the bulk of the grid and at the
boundary $y=y_{max}$. Note, however, that this effect is not as dramatic
as in the case of HR and the Hamiltonian constraint violation reduction
(compare with figures 1 and 2 of Paper I). Figure \ref{J_Short_Runs} shows
the evolution of the total angular momentum of the BNS normalized to its
initial value. Note that the use of MR (thick solid line) improves the
quality of the run only marginally over the performance obtained with
HR alone (thin solid line). The small grid size for which and orbital
period is roughly equivalent to 10 side-to-side light crossing times
makes this result quite remarkable. The BSSN simulation is presented with
a dotted line.

\subsection{Long High-Resolution Large-Grid Runs}
\label{LRC}

\begin{figure}
\begin{center}
\caption{ \underline{Long Runs}: Evolution of the $L_2$ norm of the
momentum constraint violation across the numerical grid. The
solid line corresponds to the MR+HR runs, while the dashed curve shows
the result of using only HR. The improvement on the satisfaction of the
momentum constraint is about a factor of 4 at the end of the simulation.}
\label{Mi_Long_Run}
    \includegraphics[width=5.0in]{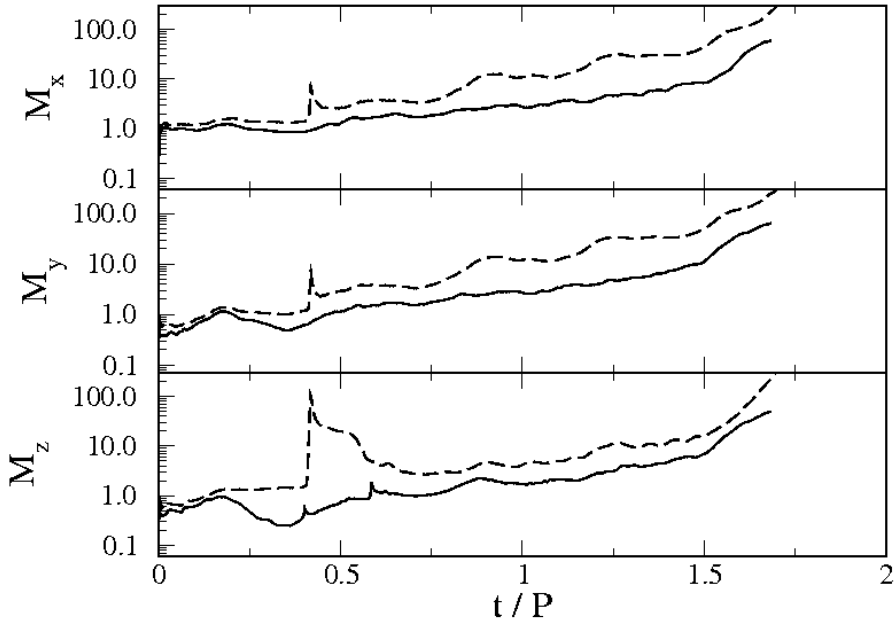}
\end{center}
\end{figure}

We tested the MR scheme with a large-grid high-resolution simulation
of an irrotational BNS system. The details of the initial data set
are provided in table 1 of Paper I and the corresponding convergence
tests are presented in \ref{appendix_CT}.  All the plots of this section
show curves that, for clarity, have been normalized to their corresponding
initial values. The MR free parameter values used for these long runs
were $\omega_M=0.01$ with 20 relaxation steps.

The evolution of the $L_2$ norm of the three components of the momentum
constraint residual $\mathcal{M}_i$ across the numerical grid is shown
in figure \ref{Mi_Long_Run}. While MR achieves the reduction of the constraint
violation as it was designed to do, the improvement
is not as impressive as in the case of HR and the Hamiltonian constraint
violation (figure 4 of Paper I). At the end of the simulation, the
momentum constraint violation was about four times smaller than in the HR
only runs. The spikes present in the HR only curves
at $t \simeq 0.4 P$ occur on the stellar surface and are related to
matter displacement in the grid, a side-effect of using a frozen
shift vector. Note, however, that those spikes disappear when using MR.

\begin{figure}
\begin{center}
\caption{ \underline{Long Runs}: Evolution of the total angular
momentum $J$. The solid, dashed, dash-dotted lines corresponds to the
MR+HR, HR only, and BSSN runs respectively. The dotted line shows the
PN estimation (see Appendix B of Paper I). The inset expands the plot for
the first half orbital period.}
\label{J_Long_Run}
    \includegraphics[width=5.0in]{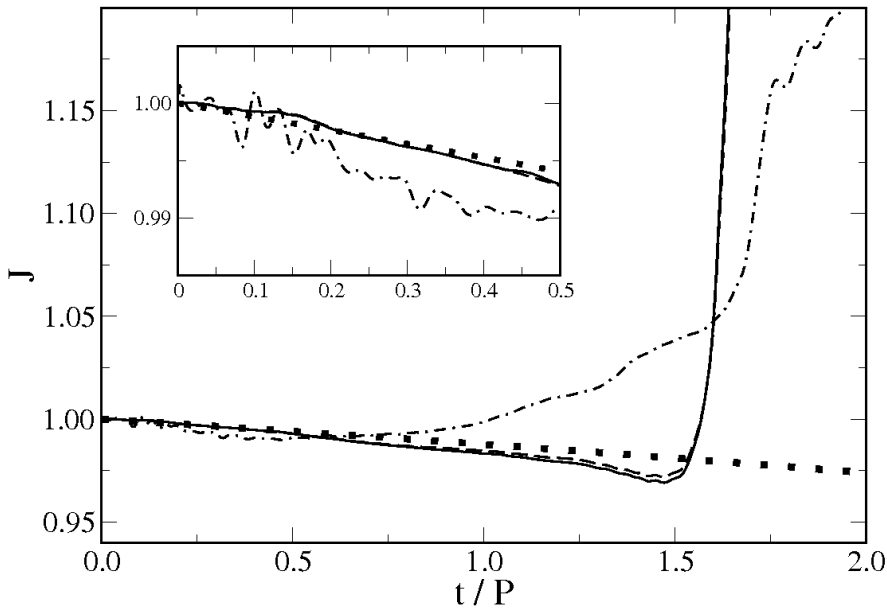}
\end{center}
\end{figure}

Figure \ref{J_Long_Run} shows the evolution of the total angular
momentum. The plot shows the MR+HR (solid), HR only (dashed) and BSSN
(dashed-dotted) curves, as well as the PN estimation (dotted line)
of the angular momentum loss for a point-mass binary with the same mass
and angular momentum as the BNS in consideration (see Appendix B of Paper I).
Both the HR only and MR+HR runs agree with the PN prediction for
about 1.5 orbital periods. The inset of figure \ref{J_Long_Run} zooms in
on the first half of the period, showing the reduced level of noise in the
HR only and MR+HR curves.

\begin{figure}
\begin{center}
\caption{ \underline{Long Runs}: Surface plot of the violation
of the $x$ component of the momentum constraint. The left (right) column
shows snapshots of the BSSN (MR+HR) run at t=0.5P, 1.0P, and 1.5P. The
lower surface plots show the rest mass density, indicating the position
of the star in the grid.}
\label{Mx_surface}
    \begin{tabular}{c|c}
    \rotatebox{-90}{\includegraphics[width=2.5in]{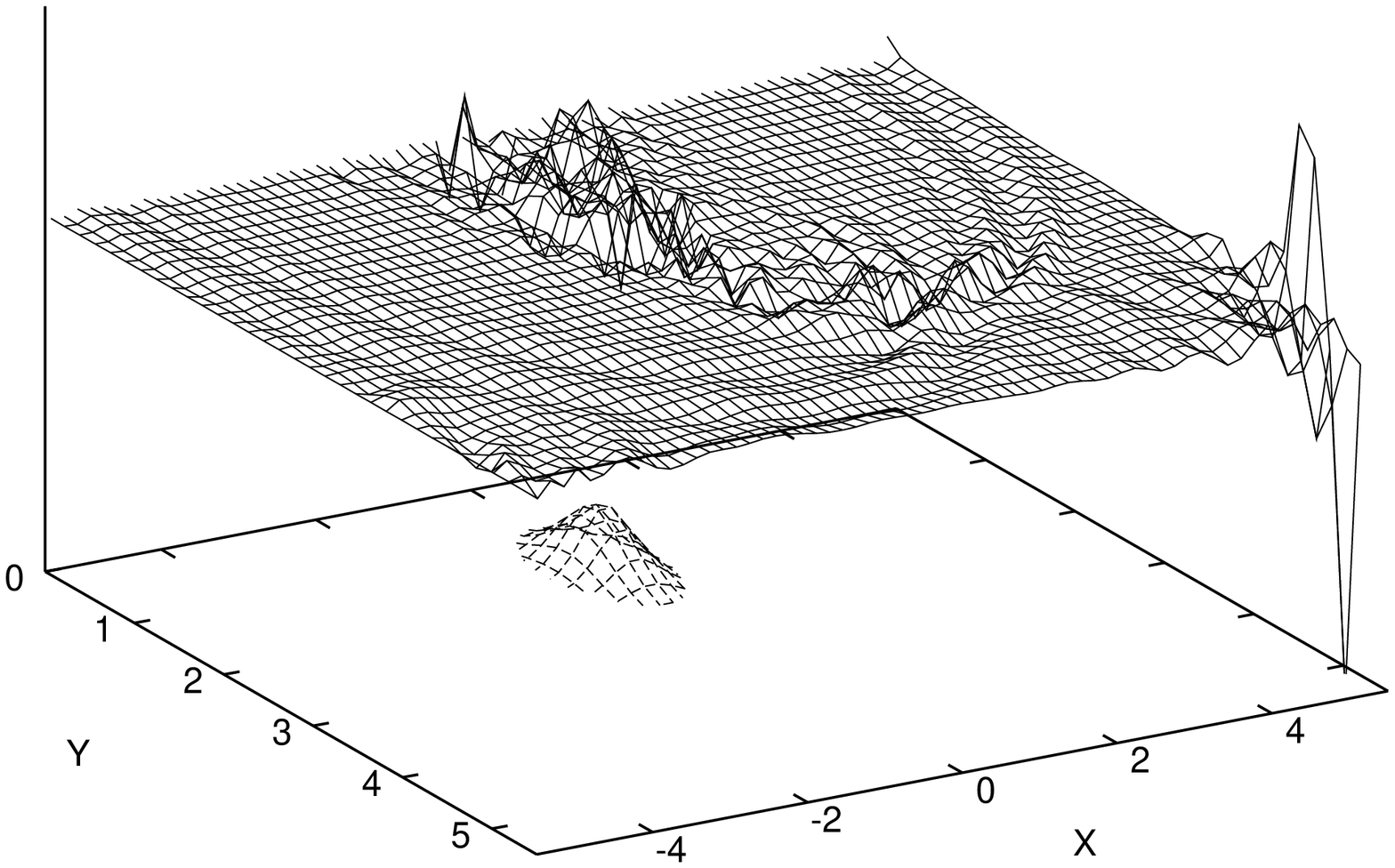}}
    \rotatebox{-90}{\includegraphics[width=2.5in]{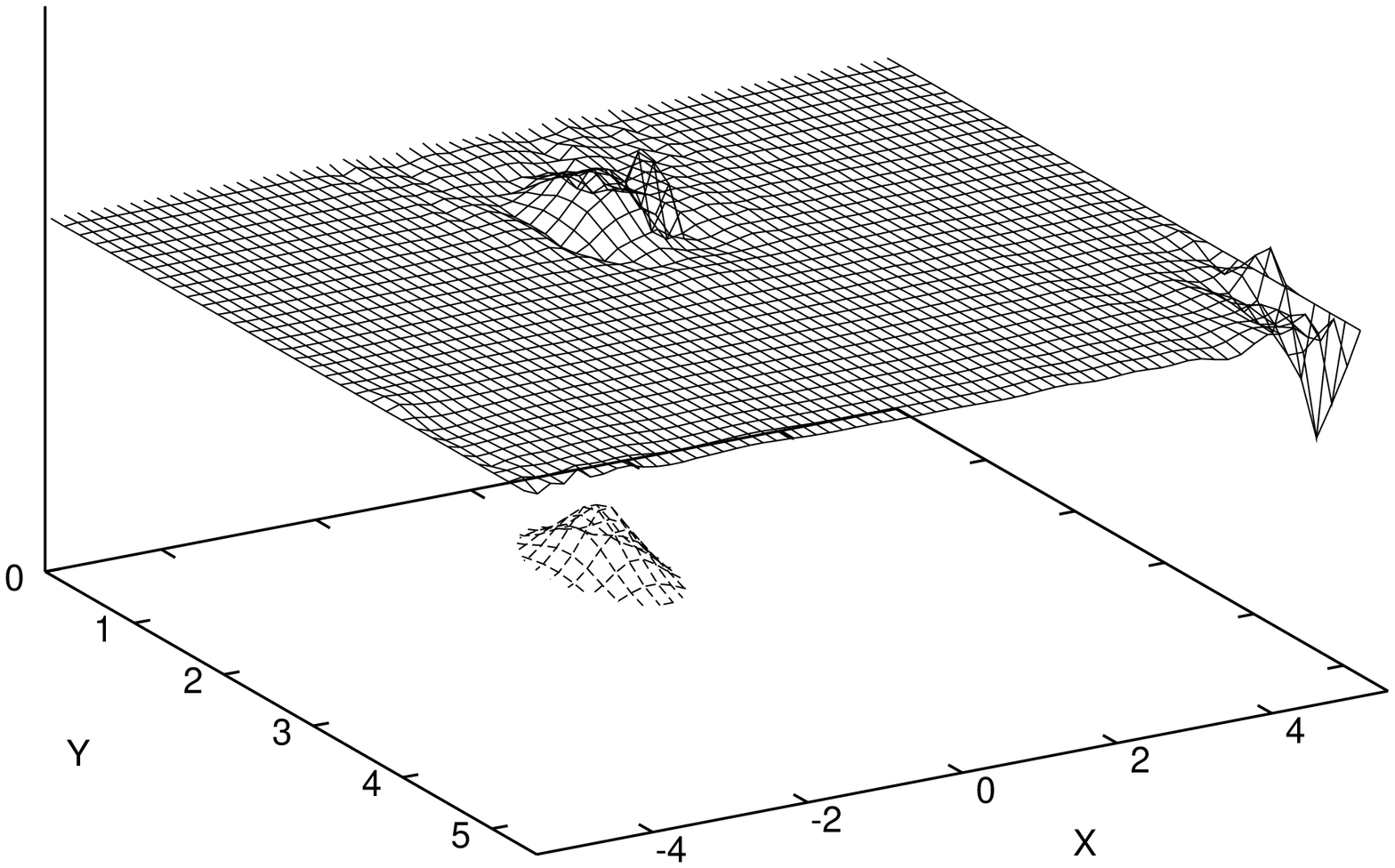}}\\
    \rotatebox{-90}{\includegraphics[width=2.5in]{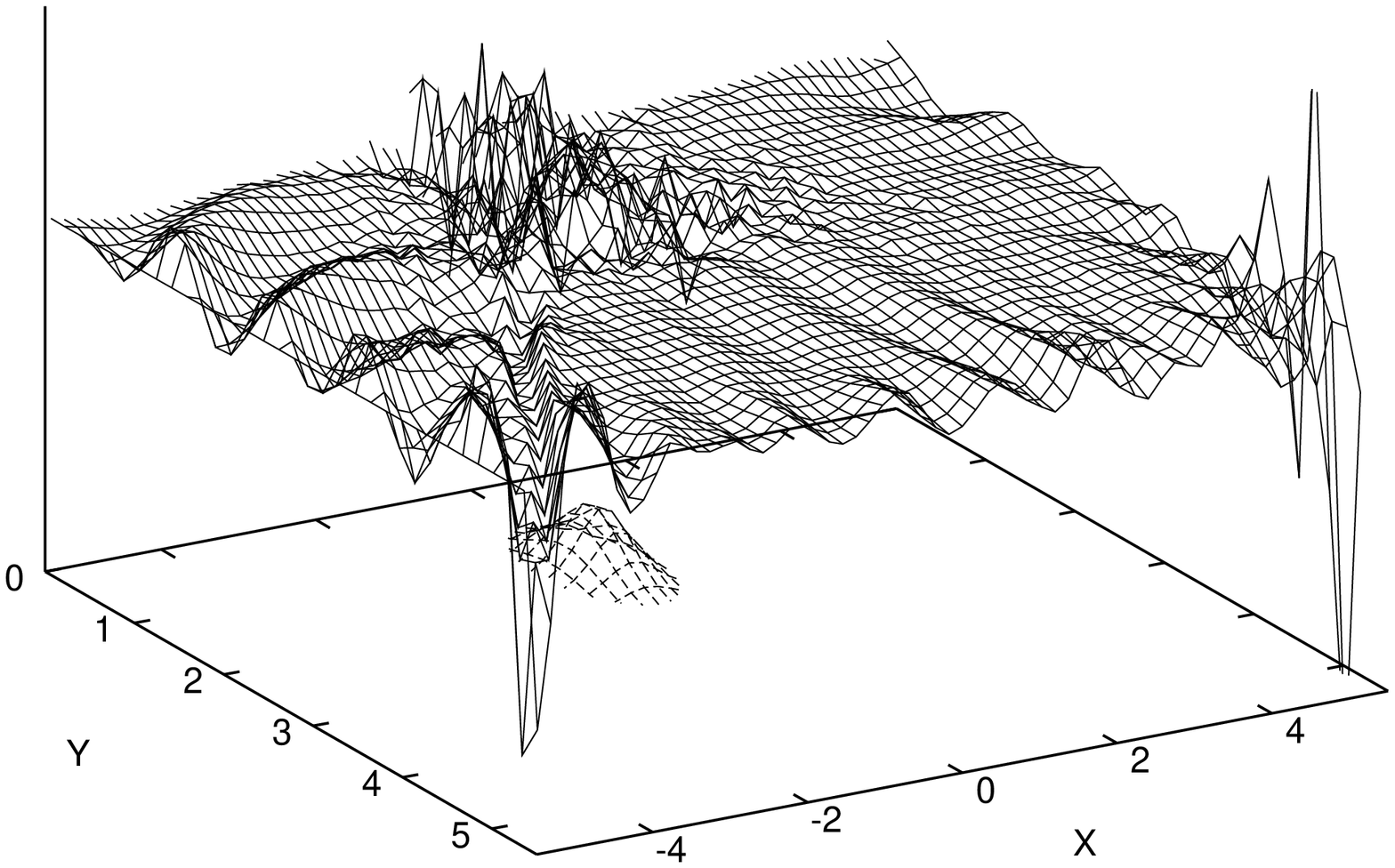}}
    \rotatebox{-90}{\includegraphics[width=2.5in]{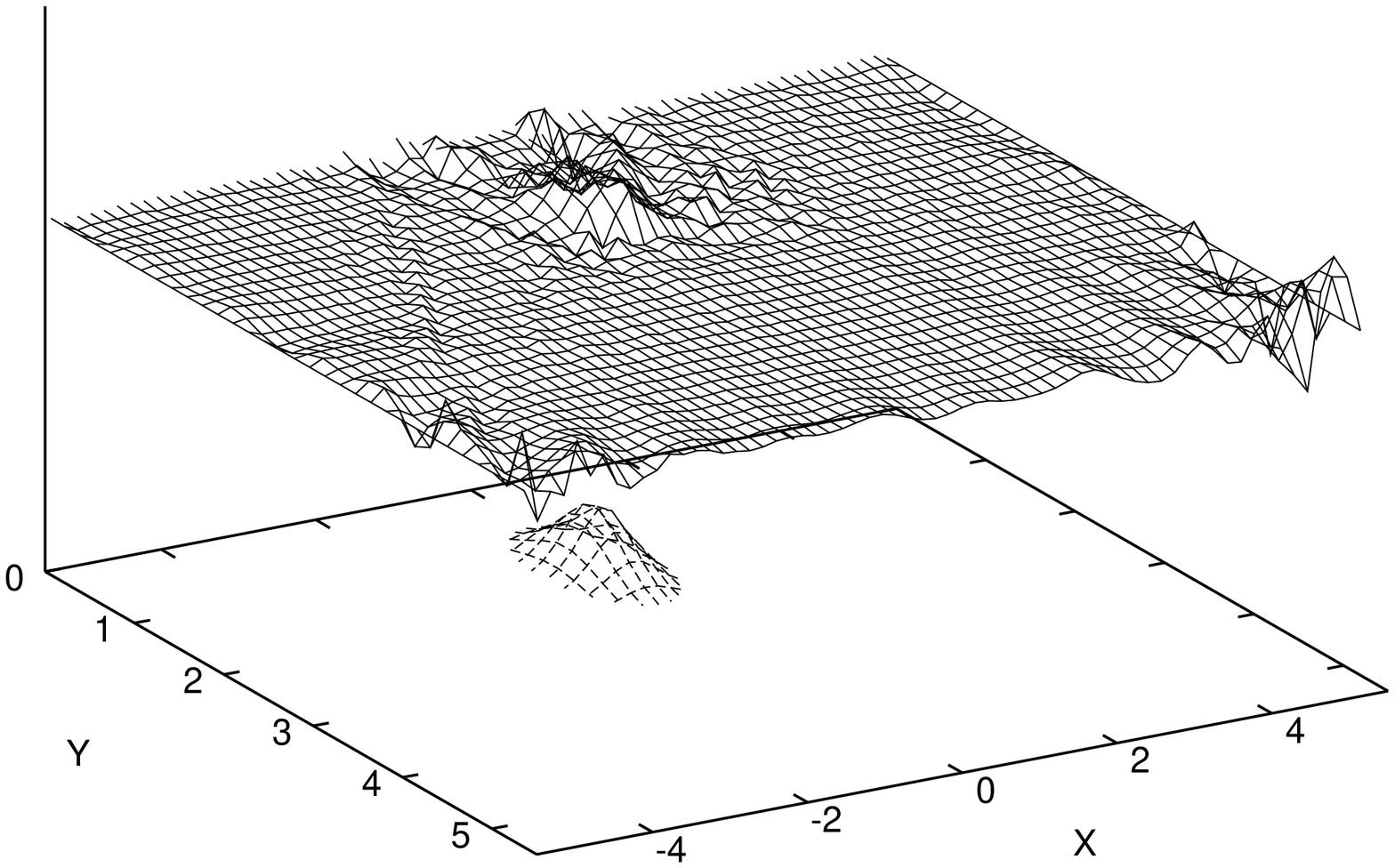}}\\
    \rotatebox{-90}{\includegraphics[width=2.5in]{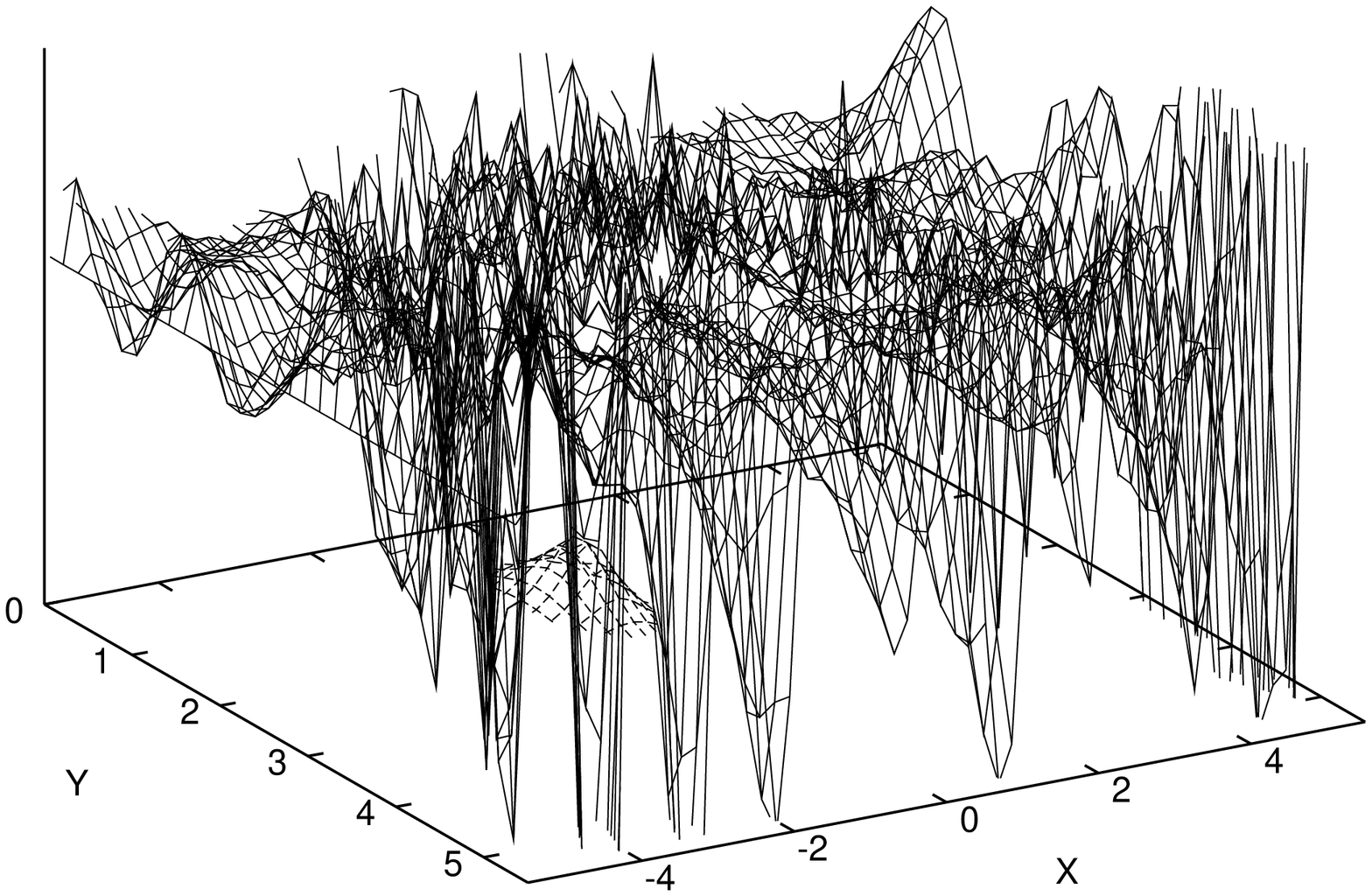}}
    \rotatebox{-90}{\includegraphics[width=2.5in]{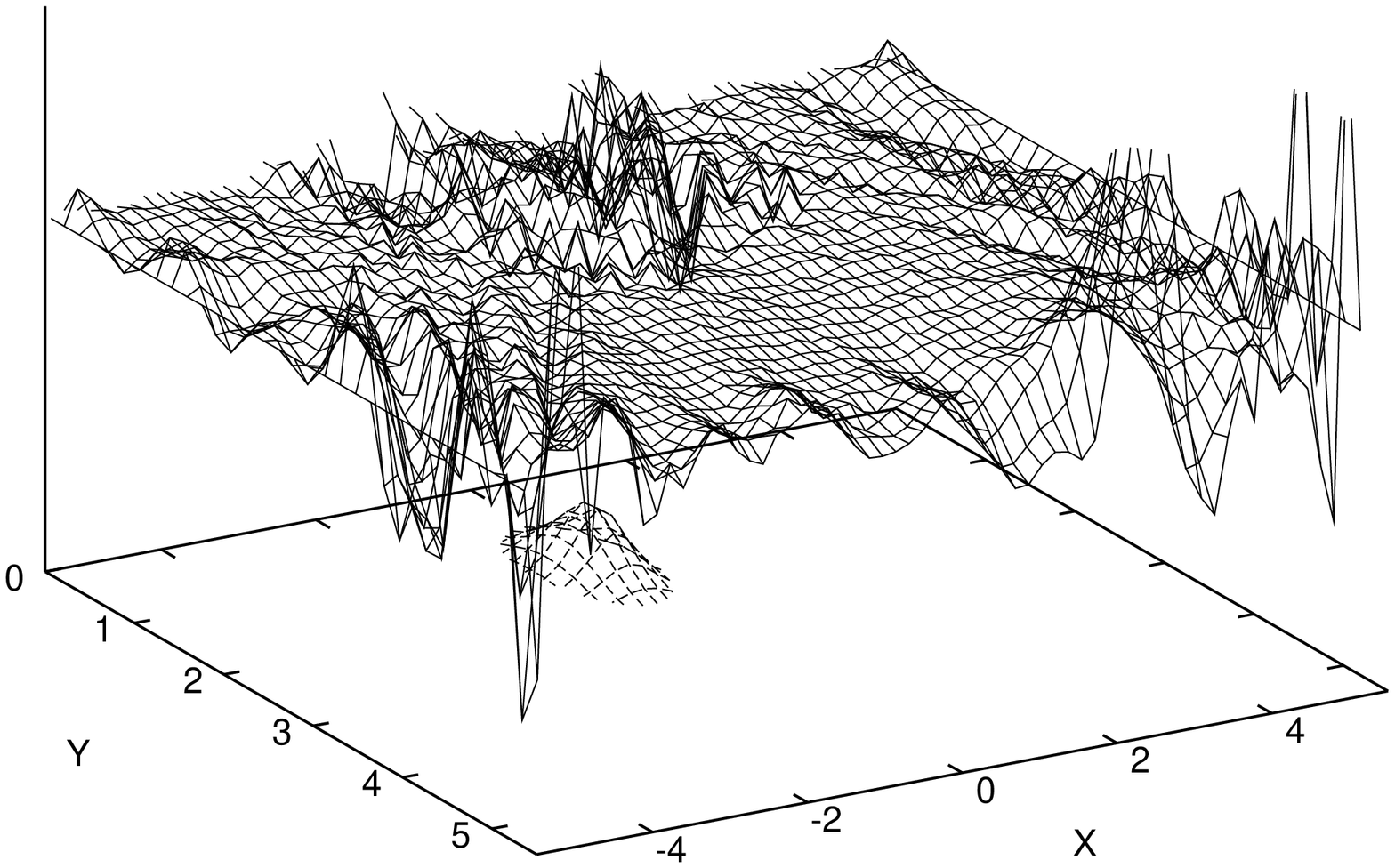}}
    \end{tabular}
\end{center}
\end{figure}

The contour plots \ref{Mx_surface} show in more detail the evolution of
the momentum constraint violation during the simulation. The left (right)
column shows snapshots of the BSSN (MR+HR) run at t=0, 0.5P, 1.0P, and
1.5P. The lower surface plots show the rest mass density, highlighting
the position of the star in the grid. The instability that eventually
stops this simulation originates at the corner of the cubical grid and
is related to the use of radiation boundary conditions in combination
with a rotating frame. While MR manages to reduce the effect of this constraint
violation, simulations based on larger grids will have to address this problem.
This effect is less serious in smaller grid runs, where the effect of frame
rotation at the corner is consequently smaller. The analysis of this instability
as well as a more comprehensive study of shift vectors for use with
constraint relaxation methods will be done in future work.

\begin{figure}
\begin{center}
\caption{ \underline{Long Runs}: Remaining quality control curves
for the MR+HR (solid), HR only (dashed), and BSSN (dash-dotted) runs.
From top to bottom, we show the evolution of the coordinate
separation between stellar centers $d$, the total gravitational $M$
mass, and the $L_2$ of the Hamiltonian constraint.}
\label{QC_Long_Run}
    \includegraphics[width=5.0in]{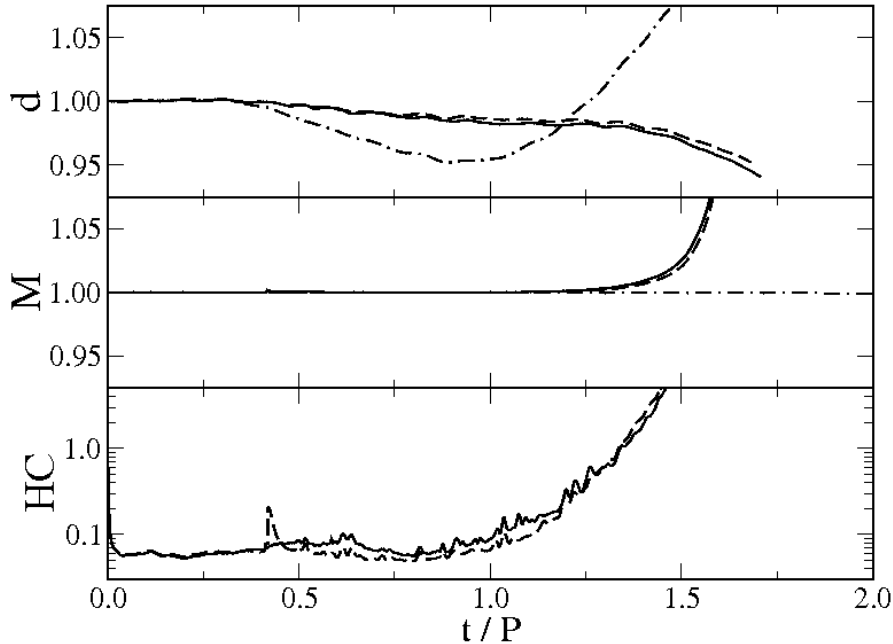}
\end{center}
\end{figure}

Figure \ref{QC_Long_Run} shows the remaining quality control
curves: the coordinate separation between stellar centers $d$, the
total gravitational $M$ mass, and the $L_2$ of the Hamiltonian constraint.
The latter curve corresponding to the BSSN run is not shown, given that
it is out of scale. The total rest mass of the system remains invariant
to within a 0.1 \% in all runs.


\section{Conclusions}
\label{conclusions}

We introduced the momentum relaxation method which
complements the Hamiltonian relaxation scheme presented
in Paper I. These algorithms gently relax the conformal factor
$\psi$ and the vector potential $w_i$ to achieve a gradual
control on the growth of constraint violating modes. This smooth
updating of the $\psi$ and $w_i$ is a key ingredient of the methods
since fully enforcing the satisfaction of the constraints (i.e.
by relaxing the fields to obtain numerical solutions of
the constraint equations) quickly renders the simulation
unstable.

The constraint relaxation methods have been tested in combination 
with BSSN in the simulation of binary neutron stars. Their use not 
only effectively quenches the constraint violating modes, but also 
improves the overall quality of the simulation as seen in the behavior 
of the total gravitational mass and angular momentum of the binary.
The simulations in this paper end after a couple
of orbits due to two principal reasons: the inadequate use of the
frozen shift condition and instabilities generated by the use
of radiation (Sommerfeld) boundary conditions for
the rest of the gravitational fields.
The boundary conditions produce constraint violating modes at the
corners of the cubical grid which grow worse with increasing grid
size. Future work will concentrate on finding the best gauge
choices as well as a way to avoid the problems generated at the corners 
of the grid. Another important aspect that remains to be studied
is the behavior of the relaxation techniques in highly dynamical
spacetimes. This will be addressed by studying their use in the
simulation of BNS mergers. Finally, it remains to be seeing how
these methods will affect the simulation in the presence of black 
holes. One potential complication could arise when using excision,
since this would require the development of inner boundary conditions
for the conformal factor and the vector potential. Note, however, 
that new techniques for black hole evolutions without excision 
have recently been developed \cite{Campanelli:2005dd,Baker:2005vv}. 
These methods have been succesfully employed in numerical simulations of 
black hole binaries
\cite{Campanelli:2005dd,Baker:2005vv,Herrmann:2006ks,Campanelli:2006gf} 
and may prove to be a more robust platform for the implementation of 
relaxation techniques.


\ack It is a pleasure to acknowledge Wolfgang Tichy for useful discussions
and Melissa Troshinsky for carefully reading the manuscript. This work
was partially supported by National Computational Science Alliance under
Grants PHY020007N. PHY050010T, and PHY050015N.


\begin{appendix}

\section{Code Tests}
\label{appendix_CT}

\begin{figure}
\begin{center}
\caption{ Convergence of the momentum relaxation results with
varying grid sizes. The convergence test is based on the long run
simulation and the plot shows the behavior of the total angular
momentum. The labels for curves are (from smallest grid to largest)
dash-dotted, dashed, and solid. All the grids have the same spatial
resolution. The dotted line shows the PN estimation.}
\label{J_MR_Conv}
    \includegraphics[width=5.0in]{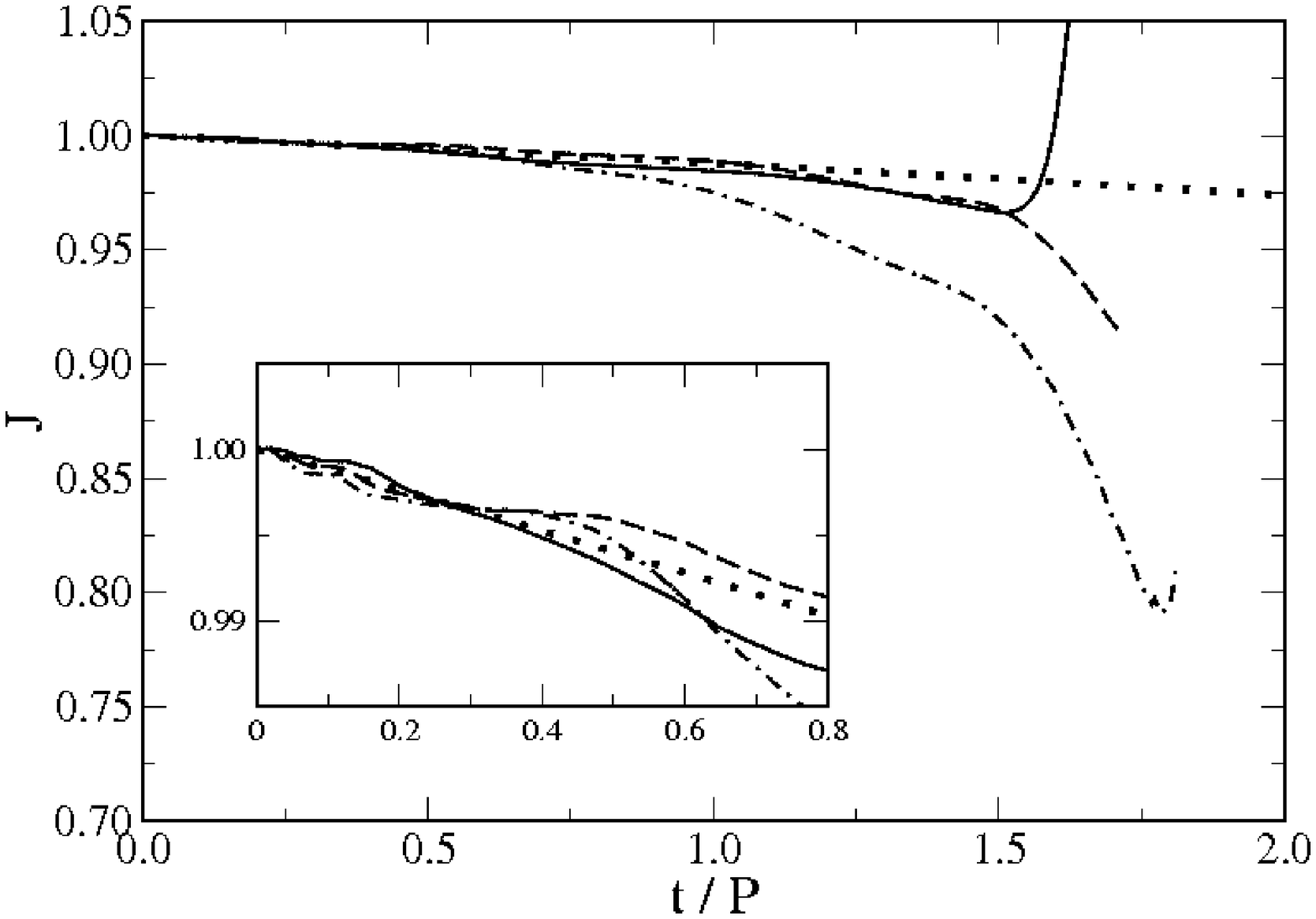}
\end{center}
\end{figure}

To test the convergence of the results obtained when using the
HR and MR, we performed the long
run simulation on three different grids with lengths (in units of total
rest mass) $=11.6$, $14.0$, and $18.6$, while keeping the same
grid spatial resolution (~ 40 grid point across the star).
Figure \ref{J_MR_Conv} shows
the evolution of the total angular momentum as a function of time.
We see the convergence of the numerical results towards the PN
estimation for point-mass binaries (dotted line). We also
notice that the corner instability occurs sooner in the largest grid
(solid line); the small grid run (dash-dotted line) stops due
to the inadequacy of $\beta$-freeze as a shift vector condition in
the presence of matter displacement.

\begin{figure}
\begin{center}
\caption{ Evolution of the $L_2$ norm of the violation of the
components of the momentum constraint for the irrotational Long Run
described in table I of Paper I with three different grid resolutions.
The dotted, dashed, and solid curves correspond to the low, medium,
and high resolution runs and the numerical factors multiplying the
curves correspond to the ratios between grid spacings.}
\label{MR_2nd_Conv}
    \includegraphics[width=5.0in]{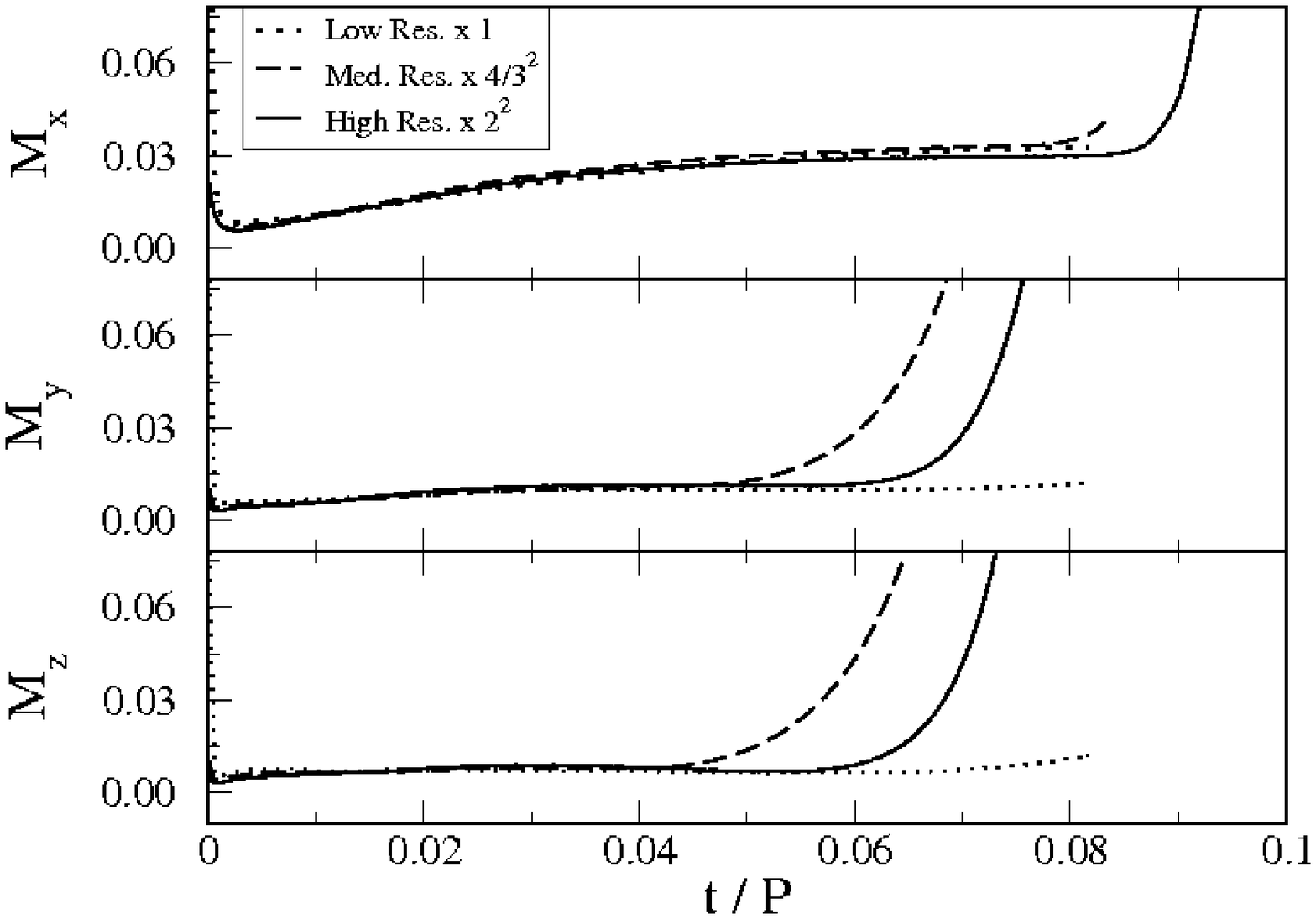}
\end{center}
\end{figure}

To test the numerical convergence of the relaxation results with
the spatial grid resolution, we performed three runs based on the 
irrotational long run using 20 (low res.), 25 (medium res.), and 40 
(high res.) grid points across the star. However, in order to see 
explicitly the second order convergence we {\it fully} relaxed the 
vector potential $w_i$ to achieve numerical solutions of the momentum 
constraint equations (\ref{MC3}). The results are shown in figure 
\ref{MR_2nd_Conv}. One important effect is that the simulations 
become quickly unstable when using full relaxation in combination 
with BSSN, the K-Driver lapse function and frozen shift vector. In 
all cases, the runs do not last beyond one tenth of an orbit. The 
reasons for this incompatibility between BSSN and the full enforcement 
of constraint satisfaction at every time step are not clear and 
deserve further study.

The Hamiltonian relaxation technique uses an equation 
derived from the Hamiltonian constraint (\ref{psi_dot}) to replace
the BSSN evolution equation for the conformal factor (\ref{cf_dot_BSSN}). 
Since the latter equation is not enforced anymore, we tested its
satisfaction by evaluating the $L_2$ norm of its violation. We define
the violation as 
\ba 
||\partial_t \Psi_{BSSN}||_2 
	&\equiv& ||\partial_t \ln(\psi) - {\mathcal{L}}_{\beta} \ln(\psi) 
		+ \frac{1}{6}\alpha K||_2 \nonumber\\~.
\ea
We implemented the finite-difference version of this equation approximating
the time derivative with a second order central difference stencil centered 
at the time step $t-1$
\ba
\fl \label{Psi_l2}
 ||\partial_t \Psi_{BSSN}||_2 \approx
    ||\frac{\ln(\psi)^t-\ln(\psi)^{t-2}}{2\Delta t} 
    - ({\mathcal{L}}_{\beta} \ln(\psi))^{t-1} 
		+ \frac{1}{6}(\alpha K)^{t-1} ||_2~.
\ea
Similarly, we tested the violation of the $xx$
component of the traceless extrinsic curvature after the vector
potential $w_i$ is updated. Calculating $\tilde A^{New}_{xx}$ as
$\tilde A^{New}_{xx} = \tilde A_{xx}+\tilde{(lw)}_{xx}$, 
we define the violation as
\ba 
\fl \label{Axx_l2}
||\partial_t \tilde A^{New}_{xx~BSSN}||_2 
	\approx  ||\frac{\tilde A^{New~t}_{xx}-\tilde A^{New~(t-2)}_{xx}}
	{2\Delta t} - ({\mathcal{L}}_{\beta} \tilde A^{New}_{xx})^{t-1} 
		- ([rhs~\tilde A^{New}_{xx}])^{t-1} ||_2~,
\ea
where $[rhs~\tilde A^{New}_{xx}]$ represents the r.h.s. of equation 
(\ref{A_dot_BSSN}).

\begin{figure}
\begin{center}
\caption{ Evolution of the $L_2$ norm of the violation of the
BSSN evolution equation for the conformal factor (top) and
the {\it xx} component of the traceless extrinsic curvature (bottom).
The dashed and solid curves correspond to the BSSN and HR+MR
short runs described in section \ref{short_runs}.}
\vskip 1cm
\label{Evol_Eqs}
    \includegraphics[width=5.0in]{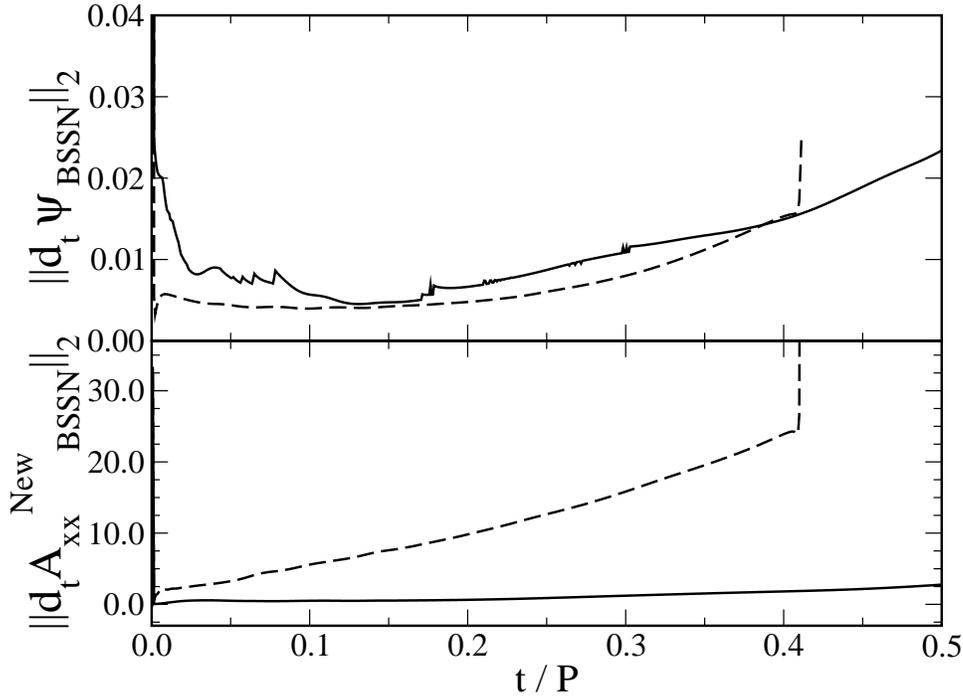}
\end{center}
\end{figure}

Figure \ref{Evol_Eqs} shows the behavior of such violations for the small
runs described in section \ref{short_runs}. The comparison of 
$||\partial_t \Psi_{BSSN}||_2$ (top) from the HR+MR run (solid curve) with the
corresponding violation from the BSSN run (dashed curve) shows no significant
difference \footnote{Note that the BSSN curve is not identically zero since 
the second order convergence in time is achieved in the evolution through the
use of ICN integration and not the time derivative stencil of equation
\ref{Psi_l2}}. The plot of $||\partial_t \tilde A^{New}_{xx~BSSN}||_2$ (bottom)
reflects a drift in the BSSN curve that is not present in the HR+MR case. This
apparent ``improvement" due to the use of relaxation methods is likely
due to the overall stability gained throughout the simulation and not a
specific betterment of the satisfaction of equation (\ref{A_dot_BSSN}).\\

\vskip 2cm

\end{appendix}


\end{document}